\documentclass[aps,prb,twocolumn,groupedaddress]{revtex4-1}

  \usepackage{amssymb}
  \usepackage{graphicx}
  \usepackage{amsmath}
  \usepackage{amsfonts}

	\usepackage[ansinew]{inputenc}
	\usepackage[usenames,dvipsnames]{pstricks}
	\usepackage{pst-grad} 
	\usepackage{pst-plot} 
	\usepackage[colorlinks,hyperindex]{hyperref}
	\hypersetup
	{
		colorlinks,%
		citecolor=black,%
		linkcolor=black,%
		urlcolor=black,%
	}

\makeindex

\begin{document}

\title{The effect of in-plane magnetic field and applied strain in quantum spin Hall systems: application to InAs/GaSb quantum wells}

\author{Lun-Hui Hu$^{1,3}$}
\author{Dong-Hui Xu$^2$}
\author{Fu-Chun Zhang$^{1,3}$}
\author{Yi Zhou$^{1,3}$}

\affiliation{$^1$Department of Physics, Zhejiang University, Hangzhou, Zhejiang, 310027, China}
\affiliation{$^2$Department of Physics, Hong Kong University of Science and Technology, Clear Water Bay, Hong Kong, China}
\affiliation{$^3$Collaborative Innovation Center of Advanced Microstructures, Nanjing 210093, China}

\date{\today}

\begin{abstract}
  Motivated by the recent discovery of quantized spin Hall effect in InAs/GaSb quantum wells\cite{du2013}\(^,\)\cite{xu2014}, we theoretically study the effects of in-plane magnetic field and strain effect to the quantization of charge conductance by using Landauer-Butikker formalism.  Our theory predicts a robustness of the conductance quantization against the magnetic field up to a very high field of 20 tesla.  We use a disordered hopping term to model the strain and show that the strain may help the quantization of the conductance.  Relevance to the experiments will be discussed.
\end{abstract}


\maketitle

\section{Introduction}

  A great deal of interest has been drawn to the discovery of quantum spin Hall (QSH) insulators, which were later identified as two dimensional (2D) time-reversal invariant topological insulators \cite{hasan2010,xlqi2011}.
  A topological insulator possesses a finite energy gap in bulk, whose low energy physics is characterized by topologically robust gapless (spin polarized) states at the edge (surface).
  Initially, such an exotic state was proposed in graphene \cite{KaneMele}, in which spin-orbit coupling is expected to open an energy gap in bulk and gives rise to counter-propagating gapless states at the edge.
  However, the spin-orbit coupling in graphene is too weak to induce an observable gap in experiment.
  The first realistic QSH insulator was theoretically proposed by Bernevig et al. \cite{bhz} in a semiconductor CdTe/HgTe/CdTe quantum well (QW), and was confirmed in edge transport experiments \cite{konig2007}.
  The second example for QSH insulator is the type II InAs/GaSb QW proposed by Liu et al. \cite{liu2008}, which has been confirmed in experiments \cite{knez2011,knez2012,suzuki2013,Mueller2015}.

  It is interesting that highly quantized longitudinal conductance plateau has been observed in lightly Si-doped InAs/GaSb QWs \cite{du2013}.
  Such a highly quantized conductance induced by dilute Si-dopants was theoretically investigated by Xu et al. \cite{xu2014}. In their theory, a single Si dopant serves as a donor or acceptor and introduces a bound state inside the bulk energy gap,
  similar to a hydrogen-like bound state in a conventional semiconductor. Their calculations show that the dilute Si-dopants are very efficient to induce in-gap localized states.

  Although the QSH system is considered to be protected by time-reversal symmetry, the edge states in InAs/GaAs QWs are quite robust against an in-plane magnetic field up to 10 tesla \cite{iknez2014,du2013} and even higher \cite{du_exciton2015}.
  Theoretically, the perpendicular magnetic field effects were studied in the context of HgTe QWs at first. It was demonstrated that HgTe QWs will undergo a QSH to quantum Hall (QH) transition as the perpendicular magnetic field increases,
  which is controlled by the band inversion \cite{gtkachov2010,jcchen2012,bscharf2012}. It was also found that weak disorders will destroy the conductance plateau in the presence of a perpendicular magnetic field \cite{maciejko2010}.
  Theoretical studies on perpendicular magnetic field effects were carried out for InAs/GaSb QWs as well \cite{pikulin2014,songbozhang2014}.
  However, the robustness of edge state against the in-plane magnetic field is not well understood.

  In this paper, we shall study the in-plane magnetic field effect in InAs/GaSb QWs. At first sight, the in-plane magnetic field will result in Zeeman splitting only and will not give rise to orbital effect in such a 2D system.
  Nevertheless, it is not ture. Becasue of the finite thickness of the InAs/GaSb interface, the orbital effect will dominate over Zeeman effect despite  the large value of Land\`{e} g-factor (as large as 10) in the system\cite{Du_communication}.
  Considering a finite spatial separation of the electron layer in InAs and the hole layer in GaSb, we can model the InAs/GaSb QW as a bilayer system. Thus, the magnetic flux penetrating between the two layers is nonzero and will naturally give rise to orbital effect.
  We shall study how such orbital effect will modify the QSH system.

  Moreover, the wide plateau in the quantized longitudinal conductance has also been observed in strained samples of InAs/GaSb QWs in the absence of Si-dopant \cite{strain}.
  It is reasonable to assume that the wide plateau in the quantized conductance in the QSH system is associated with the in-gap localized states, similar to the case in QH effect.
  In the previous theoretical studies, numerical calculation\cite{jianli2009} showed that strong Anderson disorder in HgTe QWs may drive an ordinary insulating state into a topologically non-trivial state, which is called topological Anderson insulator (TAI).
  This finding was confirmed by independent numerical simulations\cite{huajiang2009}, and may be understood within the self-consistent Born approximation\cite{cwgroth2009}.
  Anderson type bond disorders were also investigated in QSH systems \cite{juntaosong2012}.
  However, previous theories have not explained the wide plateau of the quantized conductance in the presence of applied strain. This motivates us to study strain effect in InAs/GaSb QWs.

  The rest of this paper is organized as follows. In Section \ref{sec:model}, we introduce an effective tight-binding model for InAs/GaSb QWs and consider two types of disorders.
  The formalism to calculate the longitudinal conductance and density of states (DOS) is introduced as well.
  In Section \ref{sec:magnetic}, we utilize a bilayer model to study in-plane magnetic field in InAs/GaSb QWs and focus on the robustness of helical edge states.
  In Section \ref{sec:strain}, we use a bond disorder to model the strain effect and study the in-gap localized states of the model.
  Finally, we give a brief conclusion in Section \ref{sec:summary}.

\section{Theoretical Model and Method}\label{sec:model}

  We begin with a pure system of bulk InAs/GaSb quantum wells in the absence of either external magnetic field or applied strain. Such a system can be well described by the effective BHZ model\cite{bhz,koenig2008},
  which involves four relevant orbital states \(\{\left\vert E+ \right\rangle,\left\vert H+ \right\rangle,\left\vert E- \right\rangle,\left\vert H- \right\rangle\}\), where E and H label electron and hole states, respectively and \(\pm\) denote pseudo-spins.
  The \(\mathbf{k}\cdot\mathbf{p}\) Hamiltonian\cite{liu2008} for the effective BHZ model can be rewritten as a tight-biding model on a square lattice as follows\cite{xu2014},

 \begin{align} \label{eq:bhzmodel}
     \mathcal{H}_0 = \sum_{i\sigma\alpha} V_{\alpha\sigma} c_{i\alpha\sigma}^\dagger c_{i\alpha\sigma} + \sum_{i\tau\sigma\alpha\beta} t_{\alpha\beta}^{\tau\sigma} c_{i\alpha\sigma}^{\dagger} c_{i+\tau\beta\sigma}
  \end{align}
  where $i$ is the site index, $\tau=\pm \hat{x}, \pm \hat{y}$ are the four unit vectors of nearest neighbor bonds, \(\sigma=\pm\) denotes the pseudo-spin, and \(\alpha,\beta=E,H\) denotes the orbital.

  Since the pseudo-spin $\sigma$ is conserved in a magnetic field $\mathcal{H}_0$, the $4\times4$ matrices in Eq.(\ref{eq:bhzmodel}) can be block diagonalized into two $2\times2$ matrices.
  For a given $\sigma$, in the sub-Hilbert space spanned by \(\{\vert E\sigma\rangle,\vert H\sigma\}\), \(V_{\alpha\sigma}\) is a diagonal matrix

  \begin{align}
     V_\sigma = \left(
                  \begin{array}{cc}
                    C-4D+M-4B & 0 \\
                    0 & C-4D-M+4B \\
                  \end{array}
                \right)
  \end{align}
  and \(t_{\alpha\beta}^{\tau\sigma}\) is given by the following matrix form
  \begin{align}\label{eq:hopping-t-bhz}
    t^{\pm\hat{x}\sigma} &= \left(
                              \begin{array}{cc}
                                D+B & \mp i\sigma A/2 \\
                                \mp i\sigma A/2 & D-B \\
                              \end{array}
                            \right) \nonumber \\
    t^{\pm\hat{y}\sigma} &= \left(
                              \begin{array}{cc}
                                D+B & \pm A/2 \\
                                \mp A/2 & D-B \\
                              \end{array}
                            \right)
  \end{align}
  where $A$, $B$, $C$, $D$ and $M$ are parameters, which determine the low energy band structure and will be given below.
  As discussed in Ref.\cite{xu2014}, although the lattice constant in InAs/GaSb is about 0.6 nm,
  we can choose a different lattice constant in the tight-binding model by properly choosing parameters $A$, $B$, $C$, $D$ and $M$,
  since it is an effective model derived from the \(\mathbf{k}\cdot\mathbf{p}\) Hamiltonian.
  Following Ref.\cite{xu2014}, we set the lattice constant as $a = 2$nm
  and choose $A = 0.0185$eV, $B = -0.165$eV, $C = 0$, $D = -0.0145$eV and $M = -0.0078e$V, where the terms representing bulk inversion asymmetry and structure inversion asymmetry have been neglected in $\mathcal{H}_0$.
  Note that the condition $0<M/2B<2$ guarantees that the system is in the topological insulating phase according to the Fu-Kane criterion\cite{ti_criterion}.

  To study conductance and the quantized plateau, we should consider disorders in the system. In this paper, we shall consider two types of disorders.
  The first one is that caused by impure atoms, for instance, Si-dopant at the interface. We shall model this type of disorder using on-site impurities as follows,
  \begin{align} \label{ham:anderson_imp}
    \mathcal{H}_{\text{site-imp}} = \sum_{i\sigma\alpha} W_i c_{i\alpha\sigma}^\dagger c_{i\alpha\sigma}
  \end{align}
  where the on-site potential $W_i$ is randomly distributed within a range of $(W_{\text{min}},W_{\text{max}})$. We shall discuss how to choose $W_i$ in Section \ref{sec:magnetic}.

\begin{figure}[htbp]
      \includegraphics[width=0.47\textwidth]{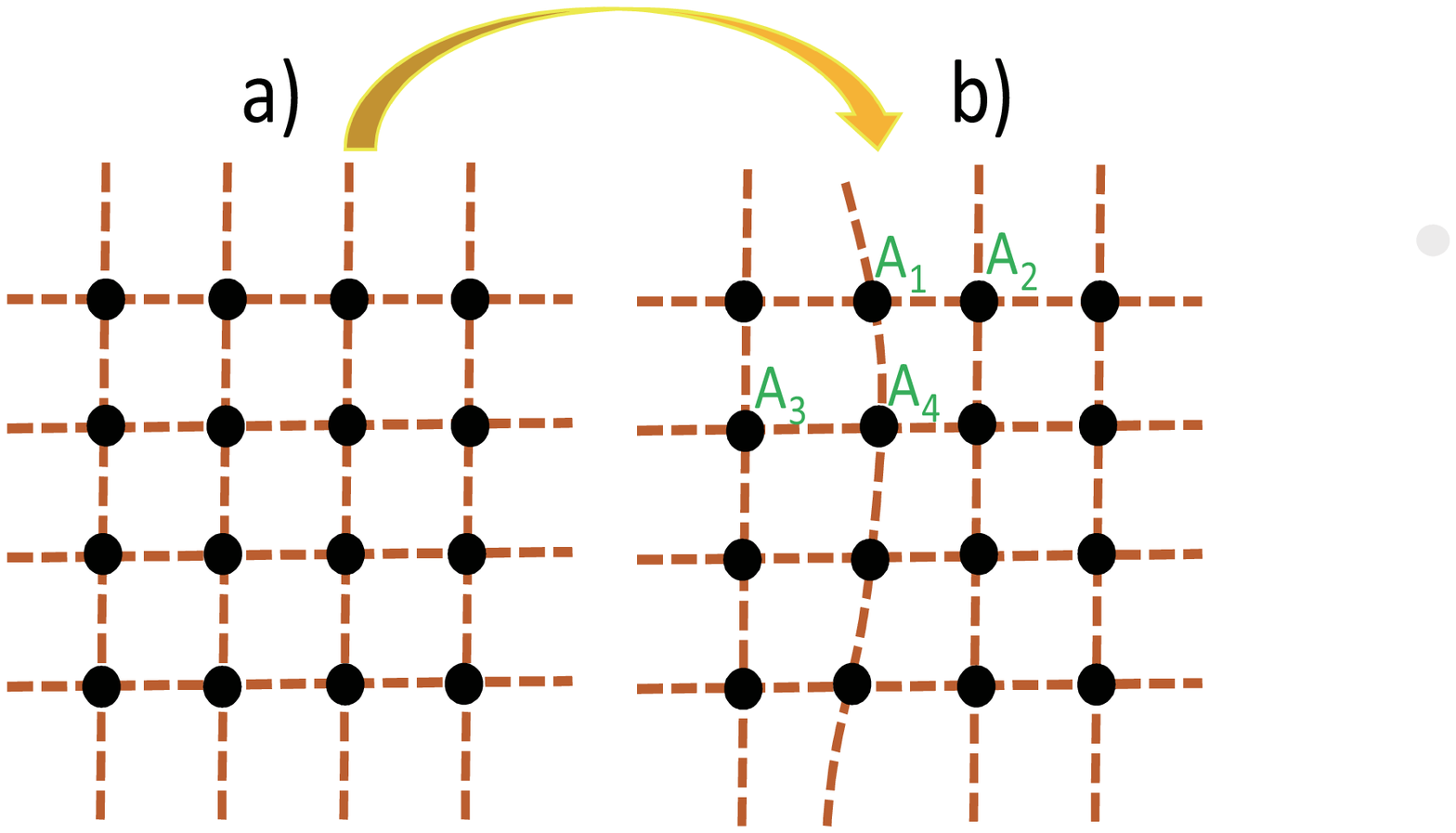}
	  \caption{Sketch figure of bond disorder. a) Undistorted lattice. b) Distorted lattice.}
	  \label{fig:bond-disorder-occur-sketch}
  \end{figure}

  The second type of disorder is the bond disorder. The InAs/GaSb quantum well is made of different crystal lattices, and we are interested in electrons and holes near the interface.
  Lattice mismatch at the interface will give rise to inherent inhomogeneity, say, lattice distortion, and stress will enhance such distortion. So we have to consider such inhomogeneity
  when the strain is applied to the system. The lattice distortion will be non-uniform even though the system is clean itself, as shown in Fig.~\ref{fig:bond-disorder-occur-sketch}.
  We model the inhomogeneous effect by bond disorder as follows,
  \begin{align} \label{ham:bond_imp}
     \mathcal{H}_{\text{bond-imp}} = \sum_{i\tau\sigma\alpha\beta} \Delta t^{i,i+\tau\sigma}_{\alpha\beta} c_{i\alpha\sigma}^{\dagger} c_{i+\tau\beta\sigma}
  \end{align}
  where the correction to the hopping integral $\Delta t^{i,i+\tau\sigma}_{\alpha\beta}$ distributes randomly within a range, which will be discussed in details in Section \ref{sec:strain}.

   In order to study transport and localization, we will calculate density of states (DOS) \(\rho\) and longitudinal conductance \(G_{xx}\) as a function of chemical potential
   using the recursive Green's function method\cite{recursive_gf} and Landauer-B\(\ddot{\text{u}}\)ttiker-Fisher-Lee formula\cite{lbfl} as in Ref.\cite{xu2014}.
   We consider a sample geometry consisting of $L_x\times L_y$ rectangular region and two semi-infinite doped metallic leads connected to the rectangle along the $x$ direction.
   (To avoid redundant scattering from mismatched interfaces between the leads and the devices, we attach two clean and doped InAs/GaSb leads as the source and drain leads.)
   This setup allows us to study DOS and the transport coefficients with both open boundary condition (OBC) and periodic boundary condition (PBC) along the $y$ direction,
   namely, PBC can be used to study bulk states and OBC can be used to study edge state transport.

   For both PBC and OPC,  the DOS can be calculated by using the Green's function $\mathcal{G}$,
  \begin{align}
     \rho = \frac{-1}{\pi L_x L_y} \text{Im} \sum_{i=1}^{L_x}\text{Tr}\,\mathcal{G}_{ii}
  \end{align}
  where $\mathcal{G}_{ii}$ is the Green's function of the $i$-th site along the $x$-direction, which can be evaluated by a recursive method\cite{recursive_gf}.

   For a generic system with the device and multiple connected terminals (leads), with the help of Landauer-B\(\ddot{\text{u}}\)ttiker formula,
   the transmission coefficients between terminal p and q are a matrix,
  \begin{align}
     T_{\text{pq}} = \text{Tr}\left\lbrack \Gamma_{\text{p}}\mathcal{G}_{\text{pq}}\Gamma_{\text{q}}\mathcal{G}_{\text{pq}}^\dagger \right\rbrack,
  \end{align}
  where \(\Gamma_{\text{p}(\text{q})}\) is calculated via \(\Gamma_{\text{p(q)}}=i[\Sigma_{\text{p(q)}}-\Sigma_{\text{p(q)}}^\dagger]\),
  and \(\Sigma_{\text{p(q)}}\) is the self-energy for lead \(\text{p(q)}\). The Green's function \(\mathcal{G}_{\text{pq}}\) between lead \(\text{p}\) and \(\text{q}\) is defined as follows,
  \begin{align}
     \mathcal{G}_{\text{pq}} = \frac{1}{E\mathbf{I}-\mathcal{H}_0-\mathcal{H}_{\text{imp}}-\Sigma_{\text{p}}-\Sigma_{\text{q}}},
  \end{align}
  where $\mathbf{I}$ is the unit matrix, and $\mathcal{H}_{\text{imp}}$ can be chosen as $\mathcal{H}_{\text{site-imp}}$ or $\mathcal{H}_{\text{bond-imp}}$ or their combination.
  To measure the longitudinal conductance, we only use two terminals in the above setup, say, $\text{p},\text{q}=1,2$.
  Thus the longitudinal conductance $G_{xx}$ can be written as
  \begin{align}
     G_{xx} = \frac{2e^2}{h} T_{12}.
  \end{align}
  It is worth noting that we can use similar setup with four terminals, say, two more leads connected to the two $y$-edges, to measure the transverse conductance $G_{xy}$ too.

  In the next sections, we shall study the effect of an external in-plane magnetic field as well as an applied strain to the transport properties.

\section{In-plane Magnetic Field Effect}\label{sec:magnetic}

  In general, a magnetic field will affect an electronic system in two ways. One is orbital effect and the other is the Zeeman splitting.
  The effect of a perpendicular magnetic field has been studied in both HgTe and InAs/GaSb quantum wells\cite{maciejko2010,pikulin2014}.
  In both cases, the perpendicular magnetic field will open an energy gap for the edge state at $\bar{\Gamma}$ point.
  To estimate this, one can choose the magnetic field $B_{\perp}=1$T.
  For HgTe quantum wells, the effective g-factor is $g_{\text{eff}}\approx20$\cite{koenig2008,netanel2011}, then the gap induced by orbital effect is
  about $3$meV, while the Zeeman effect is about $0.3$meV. It means that the orbital effect will dominate over the Zeeman spliting,
  and the later can be neglected in the analysis. For InAs/GaSb quantum wells, $g_{\text{eff}}\approx10$\cite{nilsson2006,g-factor}, the orbital effect will still
  dominate in the presence of a perpendicular magnetic field\cite{pikulin2014}.

  In the absence of external magnetic field, there are helical edge states. Both left-going and right-going channels will contribute to the quantized longitudinal conductance, resulting in $G_{xx}=2e^2/h$.
  When the perpendicular magnetic field is strong enough, the left-going (or right-going) branch will merge with the bulk states, and the right-going (or left-going) branch will become chiral edge states.
  Then the longitudinal conductance will become $e^2/h$\cite{pikulin2014}. Moreover, it was argued that the magnetic field will break time-reversal symmetry,
  thus the helical edge states are no longer symmetry-protected, disorders may localize the edge states and destroy the quantized conductance plateau\cite{maciejko2010}.

  We shall study in-plane magnetic field effect in this paper.
  At first glance, an in-plane magnetic field will cause Zeeman splitting mainly and the orbital effect can be neglected,
  since quantum spin Hall systems are quasi-two-dimensional. However, this simple judgment is not correct for InAs/GaSb quantum wells,
  which have a unique double-layered structure with electron and hole gases separated in two layers\cite{fanmingqu2015}.
  This small but finite separation (about $2$nm)\cite{Du_communication} will induce sizable orbital effect, which is much larger than the Zeeman splitting.
  So that we shall neglect the Zeeman splitting at first and examine its small correction finally.

  \begin{figure}[htbp]
		\includegraphics[width=0.47\textwidth]{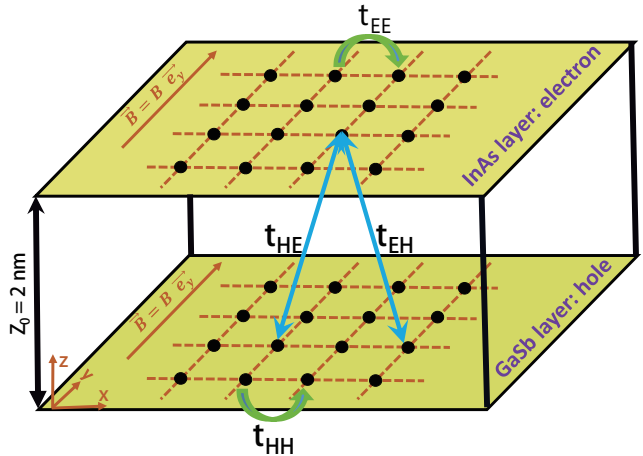}
		\caption{Bilayer system: electrons in InAs layer and holes in GaSb layer. We set \(\vec{B}=B_{\parallel}\vec{e}_y\) and choose gauge \(\vec{A}=B_{\parallel}z\vec{e}_x\),
		and also set the effective width of such a bilayer system as \(Z_0=2\text{ nm}\).}
		\label{fig:bilayer-in-plane-magnetic-field}
  \end{figure}

  To model the separation of electron and hole gases, we utilize a bilayer square lattice illustrated in Fig.~\ref{fig:bilayer-in-plane-magnetic-field}.
  In this system, there are two atomic orbitals ${\vert E \pm\rangle}$ located on the top InAs layer and two other orbitals ${\vert E \pm\rangle}$ located on the bottom GaSb layer.
  In the absence of an external magnetic field, the system is governed by the BHZ Hamiltonian $\mathcal{H}_0$ given by Eq.(\ref{eq:bhzmodel}).
  The orbital effect of an external magnetic field can be introduced by the Peierls substitution of hopping integrals $t^{ij}$ between site $i$ and $j$,
  \begin{align}
     t^{ij} \to  t^{ij} \exp\left( \frac{2\pi i}{\Phi_0}\int_{i}^{j} d\mathbf{l}\cdot \mathbf{A} \right)
  \end{align}
  where $\Phi_0=h/e$ is the magnetic flux quantum, and $t^{ij}$ refer to the hopping matrix $t^{\tau\sigma}_{\alpha\beta}$ defined in Eq.(\ref{eq:hopping-t-bhz}).

  Consider an in-plane magnetic field along the $y$-direction as shown in Fig.~\ref{fig:bilayer-in-plane-magnetic-field}, $\mathbf{B}_{\mathbf{\parallel}}=B_{\parallel}\vec{e}_y$,
  choose a Landau gauge $\mathbf{A}=B_{\parallel} z\vec{e}_x$, and set a middle point between the two layers as the origin. Then the Peierls substitution can be written explicitly
  for all the nearest neighboring bonds,
  \begin{eqnarray}
     \begin{cases}
        t^{\pm\hat{x}\sigma}_{EE} \to t^{\pm\hat{x}\sigma}_{EE}\times \exp ( \pm i\frac{2\pi}{\Phi_0} \frac{Z_0}{2}B_{\parallel} a ), \\
        t^{\pm\hat{x}\sigma}_{HH} \to t^{\pm\hat{x}\sigma}_{HH}\times \exp ( \mp i\frac{2\pi}{\Phi_0} \frac{Z_0}{2}B_{\parallel} a ), \\
        t^{\tau\sigma}_{\alpha\beta} \to t^{\tau\sigma}_{\alpha\beta}, \, \mbox{ for others},
     \end{cases}
  \end{eqnarray}
  where $Z_0$ is the separation between the two layers and will be chosen as $Z_0=2$nm in this paper.
  Note that the Hamiltonian $\mathcal{H}_0$ keeps translational invariant by this gauge choice, and the lattice momentum $k$ is still a good quantum number.
  The $\mathbf{k}$ component of $\mathcal{H}_0$ reads
  \begin{align}\label{eq:in-plane-b-bhz-k-space}
     \mathcal{H}_0(\mathbf{k}) = \left(
                                   \begin{array}{cc}
                                     \epsilon_e(\mathbf{k}) & A(k_x+ik_y) \\
                                     A(k_x-ik_y) & \epsilon_h(\mathbf{k}) \\
                                   \end{array}
                                 \right)
  \end{align}
  where $\epsilon_e=M-(B+D)(\mathbf{k}-\mathbf{k_M})^2 a^2$, $\epsilon_h=-M-(D-B)(\mathbf{k}+\mathbf{k_M})^2 a^2$,
  and we have defined the magnetic momentum shift $\mathbf{k_M}=2\pi\frac{B_{\parallel} Z_0}{\Phi_0}\hat{e}_{x}$.

  \begin{figure}[htbp]
     \includegraphics[width=0.47\textwidth]{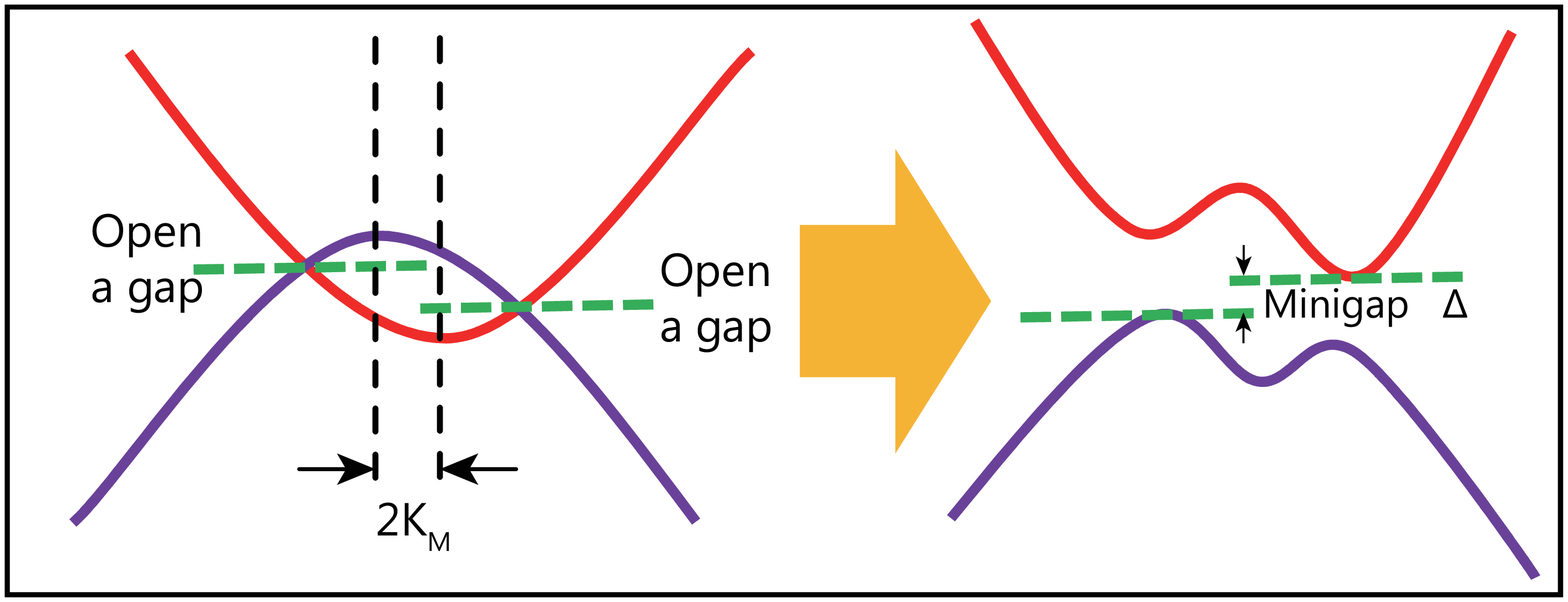}
     \caption{Hybridyzation between electron and hole band will open a ``minigap". The in-plane magnetic field will shift the electron and hole band by $\pm\mathbf{k_M}$ in $\mathbf{k}$-space respectively, and change the minigap.}
     \label{fig:sketch-minigap-openning-gap}
  \end{figure}

  Diagonalization of $\mathcal{H}_0(\mathbf{k})$ will give rise to the bulk energy dispersion and determine the energy band gap, or the ``mini-gap", as illustrated in Fig.~\ref{fig:sketch-minigap-openning-gap}.
  The in-plane magnetic field will shift the electron and hole band by $\pm\mathbf{k_M}$ in $\mathbf{k}$-space respectively.
  We find that the minigap $\Delta$ depends on the magnitude of the magnetic shift $k_M=|\mathbf{k_M}|$ linearly,
  \begin{align}
   \Delta = 2\vert A \vert \sqrt{\frac{M}{B}} \left( 1+ 2 \frac{B^2-D^2}{B \vert A \vert} k_M \right),
  \end{align}
  as long as
  $$
  \left[ 1- \left(\frac{D}{B}\right)^2\right]k_M^2 a^2 \frac{B}{M} \ll 1,
  $$
  which will be satisfied when the external magnetic field $B_{\parallel}\ll 100$T with the parameters $M,B,D$ used in this paper.
  Therefore, as the in-plane magnetic field $B_{\parallel}$ increases, the minigap $\Delta$ will close at a critical magnetic field $B_c$,
  where the electron band begin to overlap with the hole band. Such a bulk semiconductor-to-semi-metal transition has been observed in InAs/GaSb quantum wells\cite{yang1997}.

\begin{figure}[htbp]
    \begin{center}
      \includegraphics[width=0.48\textwidth]{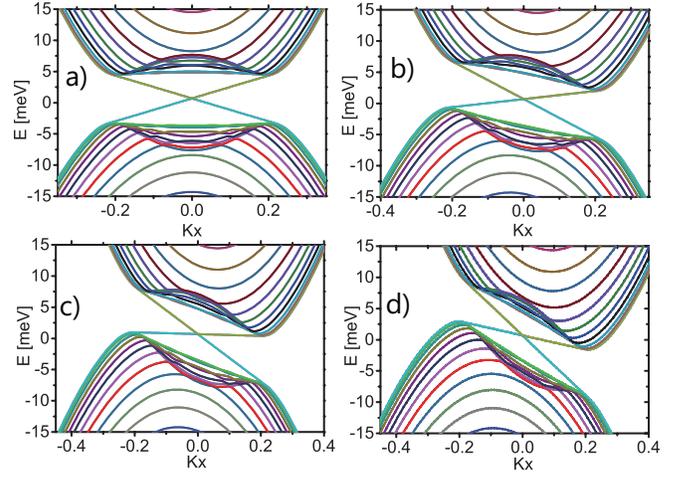}
      \caption{ The evolutoin of bulk and edge states in the presence of in-plane magnetic field. A semiconductor-to-semi-metal transition will occur at $B_c\approx 20$T due to band overlap.
      a) \(B_{\parallel}=0\,T\). b) \(B_{\parallel}=12\,T\). c) \(B_{\parallel}=20\,T\). d) \(B_{\parallel}=30\,T\).}
      \label{fig:dispersion}
    \end{center}
  \end{figure}

  Now we shall study the edge states. As shown in Fig.~\ref{fig:dispersion}, the bottom of bulk electron band will shift down and the top of hole band will shift up
  on the opposite site in $\mathbf{k}$-space when the external in-plane magnetic field is applied.
  When $B_{\parallel}>B_c\approx 20$T, a bulk semiconductor-to-semi-metal transition will happen due to the band overlap.
  It is interesting that the helical edge states survive even though the electron band overlap with the hole band.
  It is different from the case of perpendicular magnetic field, in which one of the two branches of helical edge states will merge with bulk states and helical edge states will become chiral edge states.

\begin{figure}[!htbp]
     \begin{center}
       \includegraphics[width=0.40\textwidth]{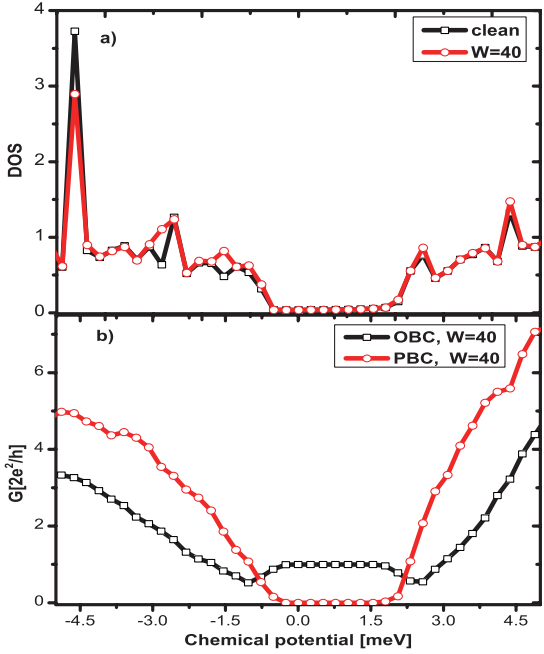}
       \caption{Bulk density of states and longitudinal conductance in the presence of Anderson disorder with strength $W=40$meV and an in-plane magnetic field $B_{\parallel}=12$T. Here we use a $L_x\times L_y=300\times 200$ strip,
       and average over 30 disordered samples.
       a) Density of States. The black line is for the clean system and the red line is for the disordered system.
       b) Longitudinal conductance for disordered systems. The black line is for the systems with open boundary condition, and the red line is for those with periodic boundary condition.}
       \label{fig:cond-obc-pbc-dos}
     \end{center}
  \end{figure}

  Then we consider how Anderson disorder will affect the edge states and induce the plateau in the quantized conductance. Anderson disorder is a type of on-site impurities
  described by Hamiltonian \eqref{ham:anderson_imp} with the on-site impurity potential $W_i$ uniformly distributed within a range of $[-W/2,W/2]$.
  Contrary to the situation of perpendicular magnetic field, where weak Anderson disorder will localize the edge states and destroy the quantized conductance plateau\cite{maciejko2010},
  the in-plane magnetic field will not affect the system so much. In the presence of an in-plane magnetic field, the Anderson disorder will neither induce in-gap localized states nor
  destroy the quantized conductance plateau. To compare it with the perpendicular field, we choose $B_{\parallel}=12$T and $W=40$meV, the later is the same value as that used in Ref.\cite{maciejko2010}.
  As plotted in Fig.~\ref{fig:cond-obc-pbc-dos}, although the in-plane magnetic field narrows the mini-gap, the Anderson disorder does not induce any additional in-gap density of states (DOS) coming from localized states.
  Thus the conductance plateau is well quantized inside the narrowed mini-gap.

\begin{figure}[!htbp]
     \begin{center}
       \includegraphics[width=0.48\textwidth]{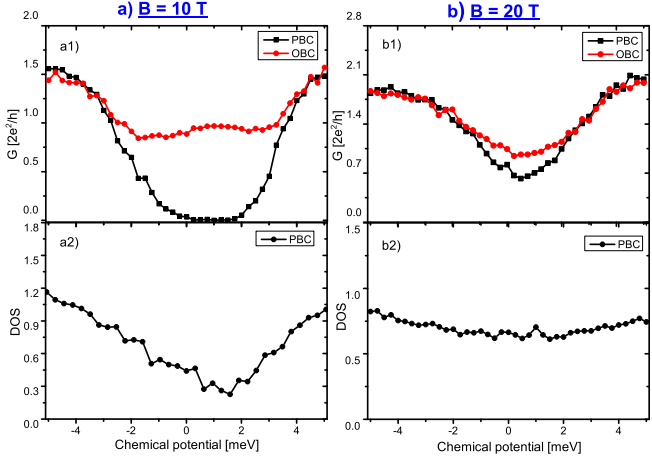}
       \caption{ Longitudinal conductance and density of states in the presence of dilute sharp disorder potential and in-plane magnetic field. We use a $400\times 200$ strip and choose 0.5\% sites randomly to be impurity sites.
       Half the impurity sites are donors and the other half are acceptors. We also set $V_{\mbox{min}}=200$meV and $V_{\mbox{max}}=300$meV, and average over 45 samples.
       a) $B_{\parallel} = 10$ T. The minigap is opened and the conductance is quantized inside the minigap. There exists finite density of states inside the minigap, indicating localized states.
       b) $B_{\parallel} = 20$ T. The minigap is closed and the conductance is not quantized.}
       \label{fig:cond-obc-pbc-dos2}
     \end{center}
  \end{figure}

  To model diluted Si-dopants in InAs/GaSb quantum wells\cite{du2013}, we should use dilute sharp impurity potential instead of usual Anderson disorder potential \cite{xu2014}.
  Namely, there are only 0.5\% sites chosen to be donors or acceptors. $W_i$ in Eq.(\ref{ham:anderson_imp}) is positive or negative for acceptor and donor sites respectively.
  The potential strength $|W_i|$ distributes randomly within a range of $(V_{\text{min}},V_{\text{max}})$.
  As shown in Fig.~\ref{fig:cond-obc-pbc-dos2}(a), similar to the system in the absence of magnetic field\cite{xu2014}, dilute sharp impurity potential will induce in-gap localized states efficiently in the presence of an in-plane magnetic field.
  Thus the conductance plateau will keep quantized inside the mini-gap as long as the gap is opened.
  However, when the increasing in-plane magnetic field drives the system from semiconductor phase to semimetal phase,
  the bulk states will contribute to the longitudinal conductance too, which will spoil the quantized conductance plateau (see Fig.~\ref{fig:cond-obc-pbc-dos2}(b)).

\section{Strain Effect} \label{sec:strain}

  As discussed in Section \ref{sec:model}, applied strain will enhance the lattice distortion due to the lattice mismatch at the interface in InAs/GaSb quantum wells.
  Such lattice distortion will modify the hopping integrals between neighboring sites and can be modeled as bond disorder.
  Indeed, bond disorder has been adopted in previous study on HgTe/CdTe quantum wells\cite{juntaosong2012}, where bond disorder is of uniformly random distribution.
  To model the lattice distortion effect, we use a bond disorder potential with different distribution by putting some randomly distributed stain sources on some lattice plaquette centers.
  A stain source located in a plaquette center $\vec{R}_p$ will contribute a Gaussian potential to each bond $t^{i,i+\tau\sigma}_{\alpha\beta}$ with a strength $V_0(\vec{R}_p)$.
  For a bond $(i,i+\tau)$ with the middle point $\vec{R}_{i,\tau}=(\vec{R}_{i}+\vec{R}_{i+\tau})/2$, we assume that the hopping integral modification due to random strain sources does not depend on spins and orbitals, and
  can be written as
  \begin{equation}\label{eq:bond-disorder-formula}
      \Delta t^{i,i+\tau\sigma}_{\alpha\beta} = \Delta t^{i,i+\tau} = \sum_{p\in G} V_0 (\vec{R}_p) \exp\left( -\frac{\vert \vec{R}_{i,\tau}-\vec{R}_{p}\vert^2}{2b^2} \right)
  \end{equation}
  where $G$ is the collect of plaquette $p$ where the Gaussian peaks locate, $b$ is the width of the Gaussian potential, and $V_0(\vec{R}_p)$ distributes uniform randomly in a range of $[-V_s,V_s]$.
  In the limit of $b\to 0$, the Gaussian potential turns to a $\delta$-function, say, a sharp potential. In this paper, we compare two situations, the sharp potential and the smooth potential with finite $b$.

  \begin{figure}[htbp]
     \begin{center}
        \includegraphics[width=0.40\textwidth]{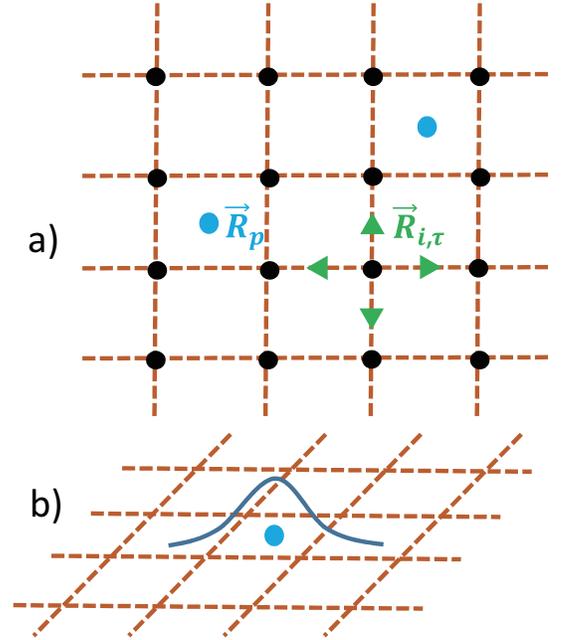}
        \caption{We consider the number of strain sources as \(0.1\times N_x\times N_y\), and these strain sources are all placed in the center of each unit cell for simplification.  }
        \label{fig:bond-disorder-smooth-sketch}
     \end{center}
  \end{figure}

\begin{figure}[htbp]
     \begin{center}
        \includegraphics[width=0.48\textwidth]{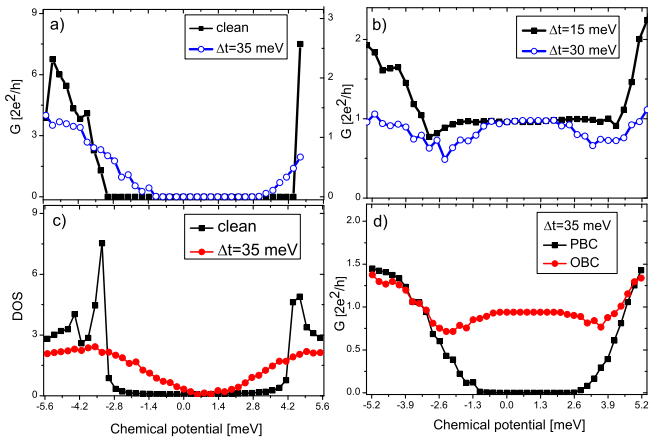}
        \caption{ Density of states and longitutdnal conductance in presence of Anderson type sharp bond disorder.
        a) Conductance from bulk states in clean and disrodered system with bond disorder strength $\Delta t=35$meV.
        b) Conductance from both bulk and edge states in disorderd systems with $\Delta t=15$meV and $\Delta t=15$meV respectively.
        c) Density of states in clean system and disordered system with $\Delta t=35$meV.
        d) Conductance from bulk states (PBC) and both bulk and edge states (OBC) in disordered system with  $\Delta t=35$meV. }
        \label{fig:bond-disorder-old}
     \end{center}
  \end{figure}

  Firstly, we study the sharp disorder potential in the limit $b\to 0$. In this limit, the bond disorder will be of Anderson type and we will simply let $\Delta t^{i,i+\tau}$ distribute randomly
  within a range of $[-\frac{\Delta t}{2},\frac{\Delta t}{2}]$ for every bond, which is equivalent to the case when $G$ covers all the plaquettes.
  For the parameters used in this paper, the hopping integral $t$ is about $180$ meV, we shall consider $\Delta t$ from $15$ meV to $35$ meV to study the bond disorder effect. The numerical results for density of stats
  and longitudinal conductance are demonstrated in Fig.~\ref{fig:bond-disorder-old}.

  By comparison of the longitudinal conductances between clean and disordered systems, see Fig.~\ref{fig:bond-disorder-old}(a), it is clear that the bond disorder will induce a mobility edge in bulk and straiten the range of gap center.
  As the disorder strength $\Delta t$ increases, the conductance plateau due to edge states will become narrower (see Fig.~\ref{fig:bond-disorder-old}(b)). However, from Fig.~\ref{fig:bond-disorder-old}(c) and (d), we find
  that, $\Delta t=35$meV is still not strong enough to induce localized states in the whole gap range. To be more efficient to localize bulk states, we consider the more realistic model with finite $b$.

\begin{figure}[htbp]
     \begin{center}
       \includegraphics[width=0.48\textwidth]{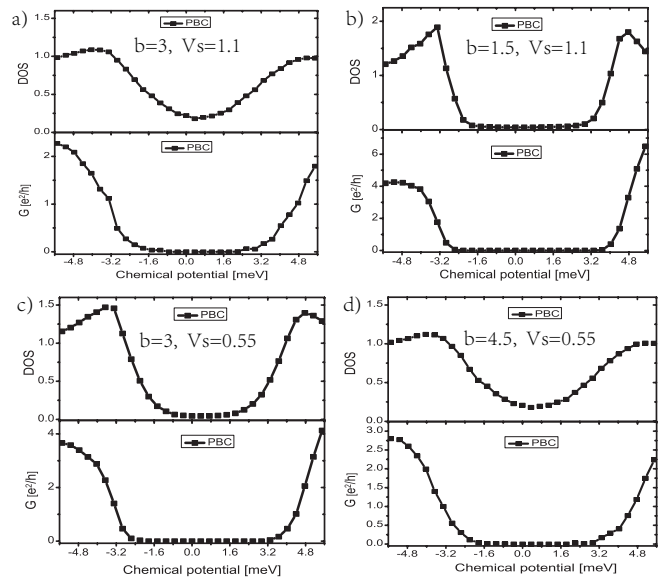}
       \caption{ Density of states and longitudinal conductance for bulk states in the presence of 10\% smooth bond disorders, with finite disorder extension $b$.
       a) $b=3$ and $V_s=1.1$meV, b) $b=1.5$ and $V_s=1.1$meV, c) $b=3$ and $V_s=0.55$meV, d) $b=4.5$ and $V_s=0.55$meV.
       }
       \label{fig:bond-disorder-smooth-cond-ods-pbc}
     \end{center}
  \end{figure}

  For the smooth potential model, we set finite $b$ and put strain sources on 10\% plaquette centers randomly. We find that the smooth disorder potential (with sufficient large $b$) has large efficiency to induce localized in-gap states in a wide range of parameters.
  For instance, when we choose $b=3$ and $V_s=1.1$meV, empirically, which gives rise to maximum $|\Delta t^{i,i+\tau}|\approx 7.5$meV corresponding to $\Delta t=15$meV in the limit $b\to 0$.
  The bond disordered potential given by this set of parameters will induce localized state in the whole gap region as shown in Fig.~\ref{fig:bond-disorder-smooth-cond-ods-pbc}(a).
  Decreasing the smooth disorder extension $b$ or the disorder strength $V_s$ will reduce the number of in-gap localized states, as we see in Fig.~\ref{fig:bond-disorder-smooth-cond-ods-pbc}(a)-(d).
  It is because the average disorder potential strength is estimate to be $(\pi b^2/5) V_s$.

\begin{figure}[htbp]
     \begin{center}
       \includegraphics[width=0.40\textwidth]{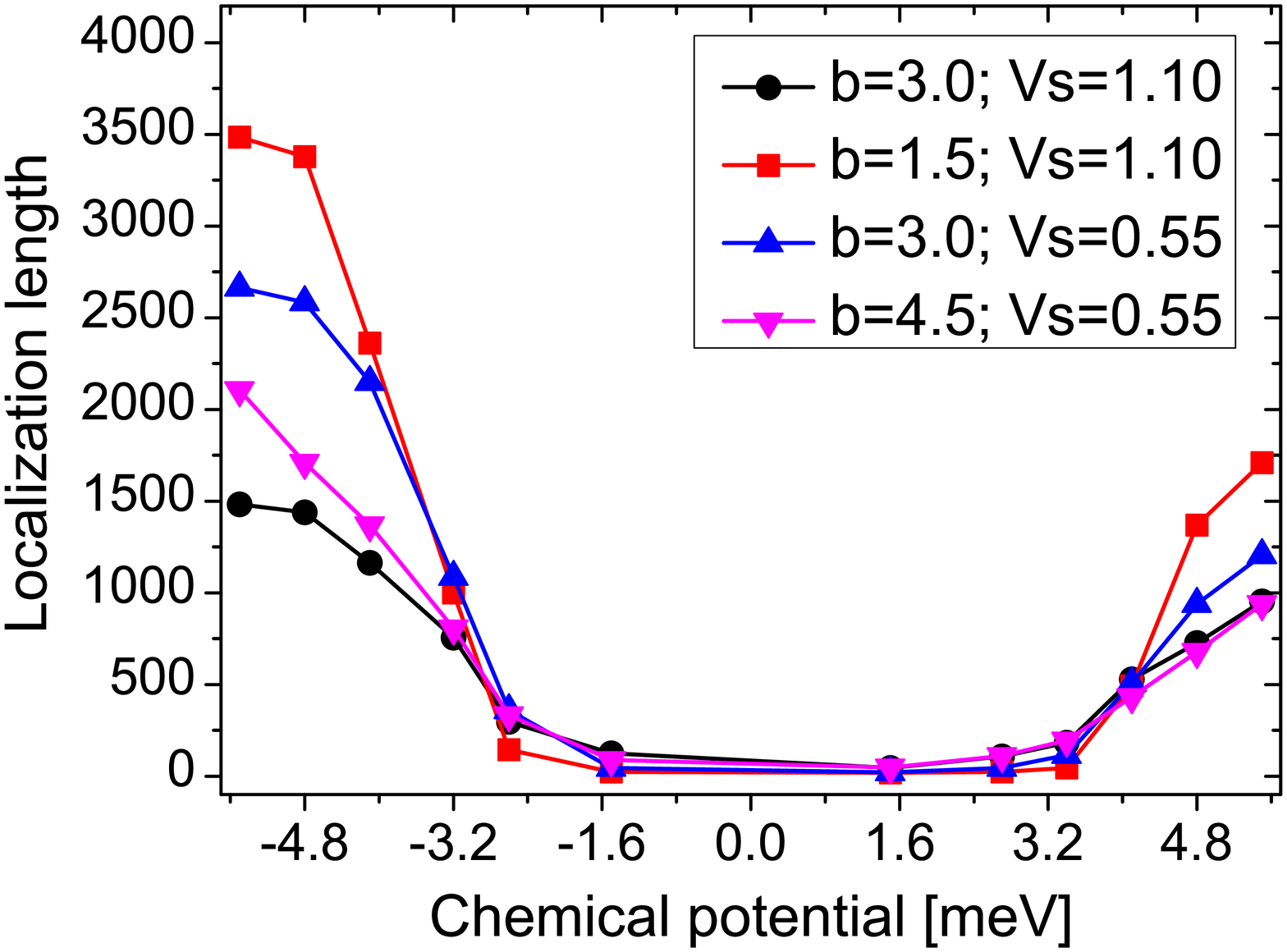}
       \caption{Localization length defined by $\xi=-2\lim_{L_{x}\to \infty}\langle\ln G/G_0\rangle^{-1}$.
       Density of states and longitudinal conductance for bulk states in the presence of 10\% smooth bond disorders, with finite disorder extension $b$ (in unit of lattice constant $a=2$nm).}
       \label{fig:bond-disorder-smooth-localization-length}
     \end{center}
  \end{figure}

  We can further examine localization lengths in such systems with finite $b$. The localization length can be extracted from the longitudinal conductance through $\xi=-2\lim_{L_{x}\to \infty}\langle\ln G/G_0\rangle^{-1}$.
  By calculating longitudinal conductance $G$ on lattices with $L_y=200$ and $L_x=300,400,500$ and averaging over 100 disorder configurations, we find out the localization length as functions of chemical potential, which is plotted in Fig.~\ref{fig:bond-disorder-smooth-localization-length}.

  By comparing the average disorder potential strength for the smooth disordered potential, $(\pi b^2/5) V_s$, with that for the sharp disorder potential, $\Delta t/4$,
  we find that the smooth bond disordered potential caused by the applied strain has much larger efficiency to induce localized in-gap states
  than sharp disordered potential in the presence of the same disorder strength.

\section{Summary} \label{sec:summary}

 In summary, we investigate in-plane magnetic field effect and strain effect in InAs/GaSb system in the parameter region with band inversion.
 For the in-plane magnetic field, we find that the orbital effect is much more significant than the Zeeman splitting due to the finite thickness of interface in the quantum well, even though the Land\`{e} g-factor is of order of 10.
 We generalized the 2D BHZ like model to a bilayer model to describe such a finite thickness effect, and study the band evolution in the presence of an tunable in-plane magnetic field.
 It is contrary to the perpendicular magnetic field, the edge states and quantized conductance plateau will persist in the presence of an in-plane magnetic field  $B_{\parallel}$ up to $B_{\parallel}\sim 20$T.
 When the in-plane magnetic field exceed 20T, the conductance quantization will become poor due to band overlap.

 We also model the applied strain effect as smooth bond disorders. Comparing with previous studied Anderson type sharp bond disorder, we find that the smooth bond disorder has much larger efficient
 to induce localized in-gap states than sharp disordered potential in the presence of the same disorder strength. This may explain why quantized conductance plateau can be observed in the experiment under applied strain,
 even though where Si dopant is absence at all.

 Finally, we note that there exist debates about the origin of the quantized conductance plateau and the observation of edge transport in the trivial insulator side \cite{Nichele2015}, as well as the identification
 of inverted or normal band gap \cite{fanmingqu2015,Nichele2015}. We hope our theoretical results can shed light on the underlying physics behind these controversial experimental observations.

\begin{acknowledgements}

We would like to thank Lingjie Du and Rui-Rui Du for help discussions. This work is partially supported by National Basic Research Program of China
(No.2014CB921201/2014CB921203), NSFC (No.11374256/11274269), and the Fundamental Research Funds for the Central Universities in China.

\end{acknowledgements}


\begin{thebibliography}{}

\bibitem{hasan2010}
  M.~Z.~Hasan and C.~L.~Kane,
  Rev.\ Mod.\ Phys.\ {\bf82}, 3045 (2010).

\bibitem{xlqi2011}
  X.-L.~Qi and S.-C.~Zhang,
  Rev.\ Mod.\ Phys.\ {\bf83}, 1057 (2011).

\bibitem{KaneMele}
 C. L. Kane and E. J. Mele, Phys. Rev. Lett. 95, 226801 (2005).

\bibitem{bhz}
  B.~A.~Bernevig, T.~L.~Hughes, and S.~C.~Zhang,
  Science {\bf314}, 1757 (2006).

\bibitem{konig2007}
  M.~K\(\ddot{\text{o}}\)nig, S.~Wiedmann, C.~Brune, A.~Roth, H.~Buhmann, L.~W.~Molenkamp, X.~L.~Qi and S. C. Zhang,
  Science {\bf318}, 766 (2007).

\bibitem{liu2008}
  C.~X.~Liu, T.~L.~Hughes, X.-L.~Qi, K.~Wang and S.-C.~Zhang,
  Phys.\ Rev.\ Lett.\ {\bf100}, 236601 (2008).

\bibitem{knez2011}
  I.~Knez, R.-R.~Du and G.Sullivan,
  Phys.\ Rev.\ Lett.\ {\bf107}, 136603 (2011).

\bibitem{knez2012}
 I.~Knez, R.~R.~Du and G. Sullivan,
 Phys.\ Rev.\ Lett.\ {\bf109},186603 (2012).

\bibitem{suzuki2013}
  K.~Suzuki, Y.~Harada, K.~Onomitsu, and K.~Muraki,
  Phys.\ Rev.\ B\ {\bf87}, 235311 (2013).
  
\bibitem{Mueller2015} S. Mueller, A. N. Pal, M. Karalic, T. Tschirky, C. Charpentier, W. Wegscheider, K. Ensslin, and T. Ihn, Phys. Rev. B 92, 081303 (2015).

\bibitem{du2013}
 L. Du, I. Knez, G. Sullivan, and R.-R. Du, Phys. Rev. Lett. 114, 096802 (2015).

\bibitem{iknez2014}
  I.~Knez, C.~T.~Rettner, S.~H.~Yang, S.~S.~P.~Parkin, L.~Du, R.~R.~Du and G.~Sullivan,
  Phys.\ Rev.\ Lett.\ {\bf112}, 026602 (2014).


\bibitem{du_exciton2015}
 Lingjie~Du, Weikai~Lou, Kai~Chang, Gerard~Sullivan, and Rui-Rui~Du,
  arXiv:1508.04509 (2015).

\bibitem{xu2014}
  Dong-Hui Xu, Jin-Hua Gao, Chao-Xing Liu, Jin-Hua Sun, Fu-Chun Zhang, Yi Zhou,
  Phys.\ Rev.\ B.\  {\bf89}, 195104 (2014).

\bibitem{gtkachov2010}
  G.~Tkachov and E.~M.~Hankiewicz,
  Phys.\ Rev.\ Lett.\ {\bf104}, 166803 (2010).

\bibitem{jcchen2012}
  J.-C.~Chen, J.~Wang and Q.-F.~Sun,
  Phys.\ Rev.\ B\ {\bf85}, 125401 (2012).

\bibitem{bscharf2012}
  B.~Scharf, A.~Matos-Abiague and J.~Fabian,
  Phys.\ Rev.\ B\ {\bf86}, 075418 (2012).

\bibitem{maciejko2010}
  Joseph~Maciejko, Xiao-Liang~Qi and Shou-Cheng Zhang,
  Phys. \ Rev.\ B\ {\bf82}, 155310 (2010).

\bibitem{pikulin2014}
  D.~I.~Pikulin, T.~Hyart, Shuo Mi, J.~Tworzyd\l{}o, M.~Wimmer and C.~W.~J.~Beenakker,
  Phys.\ Rev.\ B.\  {\bf89}, 161403(R) (2014).

\bibitem{songbozhang2014}
  Song-Bo~Zhang, Yan-Yang~Zhang and Shun-Qing~Shen,
  Phys. \ Rev.\ B\ {\bf90}, 115305 (2014).

\bibitem{Du_communication}
 Private communications with Lingjie Du and Rui-Rui Du.

\bibitem{strain}
 R.-R. Du et al., unpublished.

\bibitem{jianli2009}
  Jian~Li, Rui-Lin~Chu, J.~K.~Jain and Shun-Qing~Shen,
  Phys.\ Rev.\ Lett. {\bf102}, 136806 (2009).

\bibitem{huajiang2009}
  Hua~Jiang, Lei~Wang, Qing-feng~Sun and X.~C.~Xie,
  Phys.\ Rev.\ B\ {\bf80}, 165316 (2009).

\bibitem{cwgroth2009}
  C.~W.~Groth, M.~Wimmer, A.~R.~Akhmerov, J.~Tworzydlo and C.~W.~J.~Beenakker,
  Phys.\ Rev.\ Lett.\ {\bf103}, 196805 (2009).

\bibitem{juntaosong2012} 
  Juntao~Song, Haiwen~Liu, Hua~Jiang, Qing-feng~Sun and X.~C.~Xie,
  Phys. \ Rev.\ B\ {\bf85}, 195125 (2012).

\bibitem{koenig2008} 
  Markus~K\(\ddot{\text{o}}\)nig, Hartmut~Buhmann, Laurens~W.~Molenkamp, Taylor~L.~Hughes, Chao-Xing~Liu, Xiao-Liang~Qi and Shou-Cheng~Zhang
  J.\ Phys.\ Soc.\ Jpn.\ {\bf77}, 031007 (2008).


\bibitem{yang1997}
  M.~J.~Yang, C.~H.~Yang, B.~R.~Bennett and B.~V.~Shanabrook,
  Phys.\ Rev.\ Lett.\  {\bf78}, 4613 (1997).

\bibitem{recursive_gf}
  A.~MacKinnon,
  Z.\ Phys.\ B-Condensed matter\ {\bf59}, 385 (1985);
  G.~Metalidis, and P.~Bruno,
  Phys.\ Rev.\ B\ {\bf72}, 235304 (2005);
  G.~Metalidis,
  {\emph {Electronic Transport in Mesoscopic Systems}} (PhD thesis, Halle, 2007).

\bibitem{lbfl}
  R.~Landauer,
  Philos.\ Mag.\ {\bf21}, 863 (1970);
  M.~B\(\ddot{\text{u}}\)ttiker,
  Phys.\ Rev.\ B\ {\bf38}, 9375 (1988);
  D.~S.~Fisher and P.~A.~Lee,
  Phys.\ Rev.\ B\ {\bf23}, 6851 (1981).

\bibitem{ti_criterion}
  L.~Fu and C.~L.~Kane,
  Phys.\ Rev.\ B\ {\bf76}, 045302 (2007).


\bibitem{fanmingqu2015}
  Fanming~Qu, Arjan~J.~A.~Beukman, Stevan~Nadj-Perge, Michael~Wimmer, Binh-Minh~Nguyen, Wei~Yi, Jacob~Thorp, Marko~Sokolich, Andrey~A.~Kiselev, Michael~J.~Manfra, Charles~M.~Marcus, Leo~P.~Kouwenhoven,
  Phys. Rev. Lett. 115, 036803 (2015).

\bibitem{nilsson2006}
  K.~Nilsson, A.~Zakharova, I.~Lapushkin, S.~T.~Yen and K.~A.~Chao,
  Phys. \ Rev.\ B\ {\bf74}, 075308 (2006).


\bibitem{netanel2011}
  Netanel~H.~Lindner1, Gil~Refael and Victor~Galitski,
  Nature \ Physics \ {\bf7}, 490-495 (2011).

\bibitem{g-factor}
Xiaoyang Mu, Gerard Sullivan, Rui-Rui Du,  Appl. Phys. Lett. 108, 012101 (2016).


\bibitem{Nichele2015} F. Nichele, H. J. Suominen, M. Kjaergaard, C. M. Marcus, E. Sajadi, J. A. Folk, F. Qu, A. J. A. Beukman, F. K. de Vries, J. van Veen, S. Nadj-Perge, L. P. Kouwenhoven, B.-M. Nguyen, A. A. Kiselev, W. Yi, M. Sokolich,
M. J. Manfra, E. M. Spanton, and K. A. Moler, arXiv:1511.01728 (2015).



\end{thebibliography}

\end{document}